\documentclass{iaus}
\usepackage{graphicx}

\title[Stellar Populations in the Center of NGC~4900] 
{Stellar Populations in the Center of the Barred Galaxy NGC~4900}

\author[Cantin et al.]  %% give here short author list %% 
{Simon Cantin$^1$, Mercedes Moll\'{a}$^2$, Carmelle Robert$^1$ \& Anne
Pellerin$^3$}
\affiliation{$^1$ Universit\'{e} Laval and Observatoire du mont
 M\'{e}gantic, Qu\'{e}bec (Canada), G1K 7P4 \break email:
 simon.cantin.1@ulaval.ca,\\[\affilskip]
$^2$ CIEMAT, Avda. Complutense 22, 28040 Madrid (Spain),\\[\affilskip],
$^3$ Space Telescope Science Institute, Baltimore (USA)
}

\pubyear{2007}
\volume{xxx}  %% insert here IAU Symposium No.
\pagerange{119--126}
\date{?? and in revised form ??}
\jname{Proceedings Title IAU Symposium}
\editors{Vazdekis et al, eds.}
\begin{document}

\maketitle

\begin{abstract}

We characterize the stellar populations in the nuclear region of the
barred spiral galaxy NGC~4900 using the integral field spectrometer
OASIS and the synthesis code LavalSB and the code from Moll\'{a} \&
Garc\'{\i}a-Vargas (2000) for the young ($< 10$ Myr) and the old
stellar populations, respectively. The high spatial resolution of the
instrument allows us to find an old population uniformely distributed
and younger regions located at the end of the galaxy bar and on each
side of a nuclear bar.

\end{abstract}

\firstsection 
\section{The Star Formation History in Barred Galaxies}

Observations by \cite{cas04} have shown that not all barred spiral
galaxies, or even only those with strong bar, display stable
structures or active star formation in their nuclei. Time delays may
be important for the observations: the bar may be too young to produce
some nuclear activity, or alternatively, too old and then has already
faded into the galaxy's background. In both cases evidences could be
present in the stellar populations morphology which would be very
useful to establish an evolutionary scenario.

The star formation in the central kpc of galaxies is hardly well
described and understood.  Can we find various generations of stars
with different metallicity following a flow of gas? And then, how may
a stellar population present in the central region be discriminated
from others? Moreover, it has been proposed that fueling processes
migth involve more localized phenomena, such as nuclear bars
(\cite{shl89}), warped nuclear disks (\cite{schi00}), or nuclear
CO/stellar and HI rings (Combe et al. 2004%\cite{com04}% ). Following
the spatial distribution and the age of young ($< 10$~Myr) stellar
populations, we can obtain information on the most recent gas motion
and structures (nuclear ring or bars) in the galaxy. On the other
hand, the old population distribution and age will provide information
about previous structures or even previous starbursts. These are clues
for possible scenarios about the history of the galaxy central region.

\begin{figure*}
\begin{center}
\includegraphics[height=0.4\textheight,angle=0]{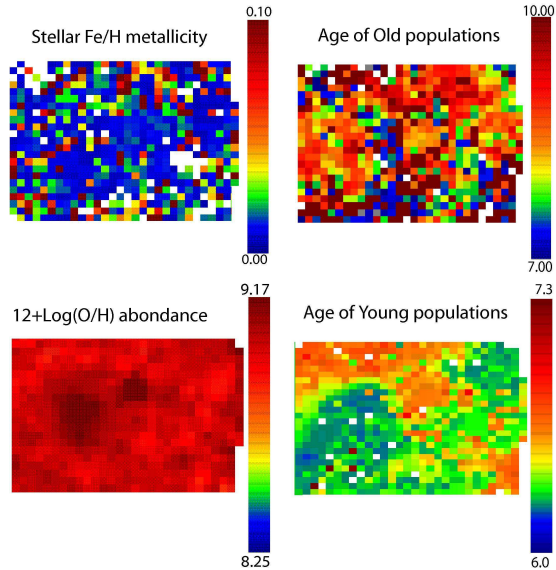}
\caption{Maps of the averaged age and metallicity for the old stellar
populations in top panels and for the younger ones at the
bottom. Average metallicity [Fe/H] and 12+log(O/H) are at the
left. Corresponding log(age) are to the right. North is up and East is
at the left.}
\label{maps}
\end{center}
\end{figure*}

\section{The Isolated SB(rs)c Galaxy NGC 4900}

Present data on the barred galaxy NGC~4900 were obtained with OASIS at
the Canada-France-Hawaii telescope in 2001. The OASIS field of view
covers the central kpc of the galaxy. The data were processed using
the XOASIS software. IRAF was then used to extract line strengths,
equivalent widths, and line ratios. Subsequent maps of extinction, ages,
and metallicities were then created to study the history of the stellar
populations.

From the absorption line equivalent widths for Mg2, Fe5270, Fe5335,
and H$\beta$ and using the code from Moll\'{a} \& Garc\'{\i}a-Vargas
(2000), it was possible to obtain the averaged age and metallicity
[Fe/H] maps for the old stellar populations (see Fig.\ref{maps}). The
first one reveals a rather uniform population of at least $\sim
100$~Myr (orange color) with a younger region of $\sim$10~Myr to the
SE. The metallicity [Fe/H] is near solar and rather uniform.  From the
equivalent width of the emission lines H$\alpha$ and H$\beta$ and
using the code LavalSB (Dionne \& Robert 2006) we estimate the age and
the oxygen abundance for the young stellar populations, too.  The age
map shows an ellongated region between 5 and 7 Myr to the SE and a
region of 8 Myr in the NW corner. These young regions are at the end
of the galactic bar on each side of a nuclear bar seen in the
extinction map (not shown here). The gas abundance is oversolar (
12+log(O/H) $\simeq$ 8.8) with a higher value for the young SE region.

In summary, we find three episodes of star formation. The first one
occurred a few 100 Myr ago involving uniformely the central region. At
the NW end of the bar, a second episode of star formation took place 8
Myr ago. The third one occurred between 5 or 6 Myr on the other side
of the nucleus. Abundance and younger star forming regions seem to
imply an inflow of high metallicity gas or recurent star formation.

\end{document}